\documentclass[]{article}

\usepackage{graphicx}
\usepackage[capposition=top]{floatrow}
\usepackage[margin=0.75in]{geometry}
\usepackage{amsmath}
\usepackage{float}
\usepackage{bm}
\usepackage{hyperref}

\title{Positivity: Identifiability and Estimability}
\author{Paul N Zivich\textsuperscript{1}, Stephen R Cole\textsuperscript{1}, Daniel Westreich\textsuperscript{1}}
\date{%
	\footnotesize
	\textsuperscript{1}Department of Epidemiology, Gillings School of Global Public Health, University of North Carolina at Chapel Hill, Chapel Hill, NC\\%
	\today
}

\begin{document}

\maketitle

\begin{abstract}
	Positivity, the assumption that every unique combination of confounding variables that occurs in a population has a non-zero probability of an action, can be further delineated as deterministic positivity and stochastic positivity. Here, we revisit this distinction, examine its relation to nonparametric identifiability and estimability, and discuss how to address violations of positivity assumptions. Finally, we relate positivity to recent interest in machine learning, as well as the limitations of data-adaptive algorithms for causal inference. Positivity may often be overlooked, but it remains important for inference.
\end{abstract}

~\\~\\

One sufficient set of identification assumptions commonly invoked are causal consistency, exchangeability, and positivity.\cite{Cole2009,Vanderweele2018,Hernan2006,Westreich2010,Greenland2017,Hernan2012,Daniel2020} Westreich and Cole (2010) previously discussed the positivity assumption and distinguished between two types.\cite{Westreich2010} Here, we reexamine deterministic and stochastic positivity, discuss how the positivity assumption can be weakened, and examine the relation of positivity to model specification. 

\section*{Definitions}
Suppose we are interested in the population average causal effect (ACE) of the action $A$ (e.g., treatment, exposure, intervention, etc.) on the outcome $Y$, where the ACE contrasts the hypothetical scenarios of everyone taking the action versus everyone not taking the action for a well-defined target population. We might write this contrast as $E[Y^1] - E[Y^0]$, where $E[\cdot]$ is the expected value function and $Y^a$ is the potential outcome under action $a$. Further, let $\Pr(\cdot)$ indicate the probability function and $\mathbf{Z}$ is a vector of covariates. 

Hereafter, attention is restricted to \textit{nonparametric} identification. As actions were not randomized by an investigator, the ACE is not identified by design. Instead, the ACE is identified under the assumptions of causal consistency, exchangeability conditional on $\mathbf{Z}$, and positivity:
\begin{equation}
	\begin{split}
		E[Y^a] = & \sum_{\mathbf{z}} E[Y^a | \mathbf{Z} = \mathbf{z}] \Pr(\mathbf{Z} = \mathbf{z}) \\
		= & \sum_{\mathbf{z}} E[Y^a | A=a, \mathbf{Z} = \mathbf{z}] \Pr(\mathbf{Z} = \mathbf{z}) \\
		= & \sum_{\mathbf{z}} E[Y | A=a, \mathbf{Z} = \mathbf{z}] \Pr(\mathbf{Z} = \mathbf{z})
	\end{split}
\end{equation}
(for continuous $\mathbf{Z}$, the summation is replaced by integration and $\Pr(\mathbf{Z} = \mathbf{z})$ is replaced with $f(\mathbf{Z} = \mathbf{z})$, where $f(\cdot)$ denotes the probability density function). Equation 1 follows from the law of total expectation, conditional exchangeability and positivity, and causal consistency, respectively. Here, attention is on the positivity assumption.

Positivity can be delineated as \textit{deterministic} positivity or \textit{stochastic} positivity.\cite{Westreich2010, Petersen2012} Deterministic positivity can be expressed as:
\begin{equation}
	\Pr(A=a | \mathbf{Z} = \mathbf{z}) \ge \epsilon > 0 \text{ for } a\in\{0,1\} \text{ and } \mathbf{z} \text{ where } p(\mathbf{z}) > 0
\end{equation}
where $p(\cdot)$ is the probability mass function (for continuous $\mathbf{Z}$, $f(\mathbf{z})$ replaces $p(\mathbf{z})$). Here, $\epsilon$ bounds the probabilities away from zero. Equation 2 is violated if there are individuals in the target population with specific combinations of variables who deterministically or structurally could not have undergone a specific action (including not receiving a particular action) or their chance is below $\epsilon$. To see why the second step in equation 1 jointly relies on exchangeability and deterministic positivity, consider a specific pattern of variables present in the target population, $\mathbf{z}'$. If those with $\mathbf{Z} = \mathbf{z}'$ could not have taken the action $A=1$, then $E[Y^1 | A=1, \mathbf{Z} = \mathbf{z}']$ is undefined. 

To provide further intuition, consider the following examples. Suppose the parameter of interest was the ACE of pregnancy on cardiovascular outcomes for all persons living in the United States. Deterministic positivity is violated here since not everyone in the defined population could become pregnant (e.g., lacking a uterus, infertile, pre- or post-menopausal, etc.) Alternatively, suppose we are interested in the ACE of an existing medication on a medical condition with a known contraindication upon admission to the hospital (e.g., penicillin and penicillin allergy).\cite{Hernan2012,Daniel2020} While not strictly deterministic (as medical mistakes happen and thus the probability is non-zero), it can be said that positivity is \textit{structurally} violated for those with the contraindication. Whether the chance of such mistakes is sufficiently likely, or such contrasts are even of practical interest, depends on substantive knowledge.

In contrast, stochastic positivity describes the probability of an action taken for a particular sample of $n$ units from the target population:
\begin{equation}
	\widehat{\Pr}_{n}(A=a | \mathbf{Z}=\mathbf{z}) > 0 \text{ for } a\in\{0,1\} \text{ and } \mathbf{z}\in\{\mathbf{z}_{1}, \mathbf{z}_{2}, ..., \mathbf{z}_{n}\}
\end{equation}
where $\widehat{\Pr}_{n}$ is
\begin{equation*}
	\widehat{\Pr}_{n}(A=a | \mathbf{Z}=\mathbf{z}) = \frac{\sum_{i=1}^{n} I(A_i =a) I(\mathbf{Z}_i = \mathbf{z})}{\sum_{i=1}^{n} I(\mathbf{Z}_i = \mathbf{z})}
\end{equation*}
Notice the hats in (3) indicate that, unlike in (2), those quantities are estimated, with the following asymmetry: if (2) does not hold then (3) cannot hold, but (3) may not hold when (2) does. In other words, whether stochastic positivity is violated depends on deterministic positivity, the possible values (dimensionality) of $\mathbf{Z}$, and the sample size $n$. Stochastic positivity violations lead to problems in estimation. To illustrate, consider the nonparametric inverse probability weighting (IPW) estimator of the ACE:
\begin{equation}
	\frac{1}{n} \sum_{i=1}^{n} \frac{Y_i A_i}{\widehat{\Pr}_n(A=1 | \mathbf{Z}_i)} - \frac{1}{n} \sum_{i=1}^{n} \frac{Y_i (1 - A_i)}{\widehat{\Pr}_n(A=0 | \mathbf{Z}_i)}
\end{equation}
If (3) is violated, (4) involves a division-by-zero and the ACE is no longer nonparametrically estimable (a similar problem occurs for nonparametric g-computation). Even if non-zero, $\widehat{\Pr}_n(A=1 | \mathbf{Z}_i)$ values close to zero can lead to instability.\cite{Khan2010,Robins2007,Cole2008,Crump2009}

Therefore, another way to frame the distinction between the types of positivity is \textit{deterministic} positivity is a problem of \textit{identifiability} and \textit{stochastic} positivity is a problem of \textit{estimability}. Here, estimability means that the parameter is (at least theoretically) possible to estimate consistently given data, and thus implies the Hadamard conditions of existence, uniqueness, and stability are met.\cite{Hadamard1902, Maclaren2019} As the stability of a solution is not implied by identifiability,\cite{Maclaren2019} identifiable parameter are not generally estimable without further assumptions (e.g., existence of a uniformly consistent estimator).\cite{Khan2010, Robins1997,Aronow2021}

\section*{Addressing Positivity Violations}

Here, we consider approaches to address positivity violations. The asymmetry between deterministic and stochastic positivity described above implies that approaches used to address deterministic positivity violations can also be used to address stochastic positivity violations, but methods to account for stochastic positivity violations may not be suitable for deterministic positivity violations. A random sample of a well-defined population is presented in Table 1, where $\mathbf{Z}=(V,W)$. As shown, one stratum of $\mathbf{Z}$ has no individuals with $A=0$. In the following, we further distinguish between two possible target populations, $S=1$ and $S=2$, from which the sample could have been drawn from. For $S=1$, (2) is violated since those with $V=0,W=0$ cannot have $A=0$. For the second population, (2) holds (i.e., $\Pr(A=0 | V=0,W=0)=0.01$ for $S=2$) and we were simply 'unlucky' in that the random sample did not include anyone with this pattern.

\subsection*{Deterministic Positivity Violations}

Deterministic positivity violations can be addressed by either redefining the target population, redefining the parameter of interest, or by opting for \textit{parametric} (or "local") identification.\cite{Petersen2012} Redefining the target population works by modifying the target population to one where deterministic positivity is no longer violated. Returning to Table 1, the target population could be redefined to consist of only those without both $V=0,W=0$ in $S=1$. The ACE for the $\mathbf{Z}$-conditional $S=1$ target population,
\begin{equation*}
	E[Y^1 | V\ne0,W\ne0] - E[Y^0 | V\ne0,W\ne0]
\end{equation*}
is then identifiable.

Instead of redefining the target population, we can also replace the ACE with another parameter.\cite{Moore2012} First, point identification of the ACE can be abandoned in favor of partial identification, like bounds.\cite{Greenland2017, Cole2019, Manski1990} However, point identification for other parameters may still be possible. Let $g^*(\mathbf{Z})$ indicate a general, investigator-specified plan that assigns actions based on $\mathbf{Z}$, where the ACE is a special case of $g^*(\mathbf{Z})$. Other plans, like $g^*(\mathbf{Z})=0$ if $V=0,W=0$ and $g^*(\mathbf{Z})=1$ otherwise, can also be considered. Specifically, this alternative $g^*(\mathbf{Z})$ avoids the deterministic positivity violation in $S=1$. For $\mathbf{Z}$-conditional plans generally, the deterministic positivity assumption is:
\begin{equation*}
	\Pr(A=a | \mathbf{Z}=\mathbf{z}) \ge \epsilon > 0 \text{ for } \mathbf{z} \text{ where } g^*(\mathbf{Z})=a
\end{equation*}
Alternatively, plans where actions are assigned probabilistically (as opposed to fixed values) could be considered instead.\cite{Munoz2012} For example, $g^*(\mathbf{Z}) = \Pr^*(A=1 | \mathbf{Z}=\mathbf{z})$ where $\Pr^*(A=1 | \mathbf{Z}=\mathbf{z})$ is an investigator-specified probability function for an action. For probabilistic plans, deterministic positivity becomes:
\begin{equation*}
	\Pr(A=1 | \mathbf{Z}=\mathbf{z}) \ge \epsilon > 0 \text{ for } \mathbf{z} \text{ where } g^*(\mathbf{Z}) > 0
\end{equation*}
Other common examples of parameters that modify or weaken he deterministic positivity assumption include the average treatment effect in the treated and population intervention effects.\cite{Westreich2014, Hubbard2008} Another alternative is incremental propensity score effects, which avoid the deterministic positivity assumption for identification.\cite{Kennedy2019, Naimi2020} However, to be informative these alternative parameters all require the existence of some $\mathbf{z}$ for which deterministic positivity is not violated. Otherwise only the natural course, the mean under the observed distribution of actions,\cite{Rudolph2022} is identified.

\subsection*{Stochastic Positivity Violations}

To reiterate, approaches to address deterministic positivity will also address stochastic positivity violations. However, some methods are unique to stochastic positivity. This section focuses on these unique methods, and therefore presumes that the data in Table 1 is a sample from the $S=2$ target population.

The most obvious solution to stochastic positivity violations is to increase $n$. As $n$ increases, stochastic positivity violations become less likely for low-dimensional $\mathbf{Z}$. However, stochastic positivity violations are only guaranteed to disappear as $n$ goes to infinity, increasing $n$ may not be practical in many settings, and simply increasing $n$ does not address stochastic positivity violations when $\mathbf{Z}$ consists of at least one continuous variable (e.g., continuous age, body mass index, etc.).\cite{Aronow2021} 

The other general technique to address stochastic positivity violations is with a statistical model. A statistical model places \textit{a priori} restriction on the set of possible distributions. These restrictions can be implemented by manipulating the data (e.g., select a subset of $\mathbf{Z}$, collapse levels of $\mathbf{Z}$ together, etc.) or by encoding the restrictions into the specified statistical model. From this model, both interpolation and extrapolation from the observed combinations of $Y,A,\mathbf{Z}$ or $A,\mathbf{Z}$ to unobserved combinations is possible. For example, we can estimate the probability of receiving $A=1$ for persons aged 55 even if no persons aged 55 had $A=1$ in the sample. One may view these model restrictions as augmentation of the data with outside information, but use of this extra-statistical information does not come freely.\cite{Rudolph2018} A further assumption is added, namely that the \textit{a priori} restrictions encoded in the statistical model are correct (i.e., the population's distribution is contained within the model's set of possible distributions). For example, consider the propensity score. The model specification assumption may be expresses as:
\begin{equation*}
	\Pr(A=a | \mathbf{Z}=\mathbf{z}) \in \mathcal{M}_\beta
\end{equation*}
where $\mathcal{M}_\beta$ is a family of probability models governed by the parameters denoted by $\beta$. Under this assumption, $\Pr(A=a | \mathbf{Z}=\mathbf{z};\beta)$ can be used to replace $\Pr(A=a | \mathbf{Z}=\mathbf{z})$. So, the model specification assumption (along with a consistent estimator of $\beta$) means equation 3 can be replaced by:
\begin{equation*}
	\Pr(A=a | \mathbf{Z}=\mathbf{z};\widehat{\beta}_n) > 0 \text{ for } a\in\{0,1\} \text{ and } \mathbf{z}\in\{\mathbf{z}_{1}, \mathbf{z}_{2}, ..., \mathbf{z}_{n}\}
\end{equation*}
Therefore, estimation of the ACE no longer requires a non-zero probability for each observed $\mathbf{z}$ but only non-zero model-estimated probabilities.

Returning to Table 1, consider the following parametric statistical model (i.e., logistic regression) to estimate the probability of $A=1$:
\begin{equation*}
	\text{logit}(\Pr(A=1 | Z)) = \beta_0 + \beta_1 V + \beta_2 W
\end{equation*}
This model imposes a restriction on the joint association (statistical interaction) of $V$ and $W$ on the probability of $A$, namely that there is none (aside: includes of the $V,W$ product term results in a saturated model, but will differ from $\widehat{\Pr}_n$ in this case because logistic models cannot return predicted probabilities of zero. However, the predicted probabilities from the saturated model will be close to zero, thus estimation problems will remain). Therefore, the ACE can be estimated with the parametric analog of equation 4 \textit{under the additional assumption that the model is correctly specified}.

A recent focus of causal inference research is the use of machine learning, or data-adaptive algorithms, to weak the model specification assumption.\cite{Chernozhukov2018,Naimi2021ML,Zivich2021,Westreich2010PS} It is important to highlight that machine learning is not a panacea for causal inference. To demonstrate, consider the use of a random forest classifier, a highly flexible machine learning algorithm, to estimate the propensity scores in Table 1 (notice: other concerns with this approach exist\cite{Chernozhukov2018,Naimi2021ML,Zivich2021}). One hyperparameter of random forests is the minimal number of observations in an endpoint, called a leaf. If the minimum leaf hyperparameter is set to 10, the estimated probabilities of $A=1$ are close to the logistic model estimates. However, setting the leaf hyperparameter to five instead can result in a division-by-zero because the random forest estimates the probability of $A=1$ as zero for those with $V=0,W=0$. This low-dimensional example illustrates that even flexible data-adaptive algorithms may require \textit{a priori} restrictions (through their hyperparameters) for estimation.

To summarize, statistical models can address \textit{stochastic} positivity violations under additional model-form assumptions. When $\mathbf{Z}$ is high-dimensional, the model specification assumption becomes a remedy to ubiquitous stochastic positivity violations. While data-adaptive algorithms may impose fewer constraints than standard parametric models, constraints become necessary in high-dimensional settings.\cite{Maclaren2019, Aronow2021}

\section*{Conclusions}

Here, we have revisited the practical implications of the positivity assumption with focus on the population inferential model,\cite{Lehmann1999} because epidemiology research is mainly concerned with populations of which only a sample is available.\cite{Lesko2017} When the $n$ units \textit{are the target population}, the definitions of deterministic and stochastic positivity coincide. While focus here was centered on positivity related to confounding, other biases and their corresponding exchangeability conditions also rely on positivity assumptions (e.g., missing data,\cite{Sun2018,Sun2018a} selection bias,\cite{Robins2000a,Howe2016} measurement error,\cite{Edwards2015} and generalizability/transportability\cite{Lesko2017, Westreich2017c}). Building on previous work,\cite{Maclaren2019,Aronow2021} we laid out how deterministic positivity relates to nonparametric identifiability and stochastic positivity to estimability. Finally, we emphasized how stochastic positivity violations can be ameliorated by additional estimability assumptions, like statistical model specification. While positivity is often overlooked, positivity remains a key assumption that should be examined by epidemiologists.

\section*{Acknowledgments}
This work was supported in part by T32-AI007001 (PNZ) and R01-AI157758 (SRC).

\newpage

\begin{table}[H]
	\caption{Illustrative example of a positivity violation\textsuperscript{*} in a random sample from a target population}
	\centering
	\begin{tabular}{llccccc}
		\hline
		&               & \multicolumn{2}{c}{$A=1$} &  & \multicolumn{2}{c}{$A=0$} \\ \cline{3-4} \cline{6-7} 
		&               & $Y=1$       & $Y=0$       &  & $Y=1$       & $Y=0$       \\ \cline{3-4} \cline{6-7} 
		\multicolumn{2}{l}{$V=1$} &             &             &  &             &             \\
		& $W=1$         & 7           & 10          &  & 18          & 32          \\
		& $W=0$         & 9           & 29          &  & 11          & 15          \\
		\multicolumn{2}{l}{$V=0$} &             &             &  &             &             \\
		& $W=1$         & 15          & 60          &  & 11          & 14          \\
		& $W=0$         & 7           & 8           &  & 0           & 0           \\ \hline
	\end{tabular}
	\floatfoot{$A$: action of interest, $Y$: outcome of interest, $V,W$: variables assumed to render the potential outcomes and action taken to be independent (e.g., exchangeable, d-separated, etc.)\\
	* The data is consistent with either a deterministic or stochastic positivity violation. Distinguishing between the types requires background knowledge or assumptions. For the target population $S=1$, the zero cells for $A=0$ are considered to result from a deterministic positivity violation. For the target population $S=2$, the zero cells for $A=0$ are deemed to be a result of a stochastic positivity violation and \textit{not} a deterministic positivity violation.}
\end{table}

\small
\bibliography{biblio}{}

\begin{thebibliography}{10}

\bibitem{Cole2009}
S.~R. Cole and C.~E. Frangakis, ``The consistency statement in causal
  inference: a definition or an assumption?,'' {\em Epidemiology (Cambridge,
  Mass.)}, vol.~20, no.~1, pp.~3--5, 2009.

\bibitem{Vanderweele2018}
T.~J. VanderWeele, ``On well-defined hypothetical interventions in the
  potential outcomes framework,'' {\em Epidemiology (Cambridge, Mass.)},
  vol.~29, no.~4, p.~e24, 2018.

\bibitem{Hernan2006}
M.~A. Hern{\'a}n and J.~M. Robins, ``Estimating causal effects from
  epidemiological data,'' {\em Journal of Epidemiology \& Community Health},
  vol.~60, no.~7, pp.~578--586, 2006.

\bibitem{Westreich2010}
D.~Westreich and S.~R. Cole, ``Invited commentary: positivity in practice,''
  {\em American Journal of Epidemiology}, vol.~171, no.~6, pp.~674--677, 2010.

\bibitem{Greenland2017}
S.~Greenland, ``For and against methodologies: some perspectives on recent
  causal and statistical inference debates,'' {\em European Journal of
  Epidemiology}, vol.~32, no.~1, pp.~3--20, 2017.

\bibitem{Hernan2012}
M.~A. Hern{\'a}n, ``Beyond exchangeability: the other conditions for causal
  inference in medical research,'' 2012.

\bibitem{Daniel2020}
R.~Daniel, ``A new template for empirical studies: From positivity to
  positivity,'' {\em Statistical Science}, vol.~35, no.~3, pp.~476--478, 2020.

\bibitem{Petersen2012}
M.~L. Petersen, K.~E. Porter, S.~Gruber, Y.~Wang, and M.~J. Van Der~Laan,
  ``Diagnosing and responding to violations in the positivity assumption,''
  {\em Statistical Methods in Medical Research}, vol.~21, no.~1, pp.~31--54,
  2012.

\bibitem{Khan2010}
S.~Khan and E.~Tamer, ``Irregular identification, support conditions, and
  inverse weight estimation,'' {\em Econometrica}, vol.~78, no.~6,
  pp.~2021--2042, 2010.

\bibitem{Robins2007}
J.~Robins, M.~Sued, Q.~Lei-Gomez, and A.~Rotnitzky, ``Comment: Performance of
  double-robust estimators when" inverse probability" weights are highly
  variable,'' {\em Statistical Science}, vol.~22, no.~4, pp.~544--559, 2007.

\bibitem{Cole2008}
S.~R. Cole and M.~A. Hern{\'a}n, ``Constructing inverse probability weights for
  marginal structural models,'' {\em American Journal of Epidemiology},
  vol.~168, no.~6, pp.~656--664, 2008.

\bibitem{Crump2009}
R.~K. Crump, V.~J. Hotz, G.~W. Imbens, and O.~A. Mitnik, ``Dealing with limited
  overlap in estimation of average treatment effects,'' {\em Biometrika},
  vol.~96, no.~1, pp.~187--199, 2009.

\bibitem{Hadamard1902}
J.~Hadamard, ``Sur les probl{\`e}mes aux d{\'e}riv{\'e}es partielles et leur
  signification physique,'' {\em Princeton University Bulletin}, pp.~49--52,
  1902.

\bibitem{Maclaren2019}
O.~J. Maclaren and R.~Nicholson, ``What can be estimated? identifiability,
  estimability, causal inference and ill-posed inverse problems,'' {\em arXiv
  preprint arXiv:1904.02826}, 2019.

\bibitem{Robins1997}
J.~M. Robins and Y.~Ritov, ``Toward a curse of dimensionality appropriate
  (coda) asymptotic theory for semi-parametric models,'' {\em Statistics in
  Medicine}, vol.~16, no.~3, pp.~285--319, 1997.

\bibitem{Aronow2021}
P.~Aronow, J.~M. Robins, T.~Saarinen, F.~S{\"a}vje, and J.~Sekhon,
  ``Nonparametric identification is not enough, but randomized controlled
  trials are,'' {\em arXiv preprint arXiv:2108.11342}, 2021.

\bibitem{Moore2012}
K.~L. Moore, R.~Neugebauer, M.~J. van~der Laan, and I.~B. Tager, ``Causal
  inference in epidemiological studies with strong confounding,'' {\em
  Statistics in Medicine}, vol.~31, no.~13, pp.~1380--1404, 2012.

\bibitem{Cole2019}
S.~R. Cole, M.~G. Hudgens, J.~K. Edwards, M.~A. Brookhart, D.~B. Richardson,
  D.~Westreich, and A.~A. Adimora, ``Nonparametric bounds for the risk
  function,'' {\em American Journal of Epidemiology}, vol.~188, no.~4,
  pp.~632--636, 2019.

\bibitem{Manski1990}
C.~F. Manski, ``Nonparametric bounds on treatment effects,'' {\em The American
  Economic Review}, vol.~80, no.~2, pp.~319--323, 1990.

\bibitem{Munoz2012}
I.~D. Mu{\~n}oz and M.~Van Der~Laan, ``Population intervention causal effects
  based on stochastic interventions,'' {\em Biometrics}, vol.~68, no.~2,
  pp.~541--549, 2012.

\bibitem{Westreich2014}
D.~Westreich, ``From exposures to population interventions: pregnancy and
  response to hiv therapy,'' {\em American Journal of Epidemiology}, vol.~179,
  no.~7, pp.~797--806, 2014.

\bibitem{Hubbard2008}
A.~E. Hubbard and M.~J. Van~der Laan, ``Population intervention models in
  causal inference,'' {\em Biometrika}, vol.~95, no.~1, pp.~35--47, 2008.

\bibitem{Kennedy2019}
E.~H. Kennedy, ``Nonparametric causal effects based on incremental propensity
  score interventions,'' {\em Journal of the American Statistical Association},
  vol.~114, no.~526, pp.~645--656, 2019.

\bibitem{Naimi2020}
A.~I. Naimi, J.~E. Rudolph, E.~H. Kennedy, A.~Cartus, S.~I. Kirkpatrick, D.~M.
  Haas, H.~Simhan, and L.~M. Bodnar, ``Incremental propensity score effects for
  time-fixed exposures,'' {\em Epidemiology (Cambridge, Mass.)}, vol.~32,
  no.~2, pp.~202--208, 2020.

\bibitem{Rudolph2022}
J.~E. Rudolph, A.~Cartus, L.~M. Bodnar, E.~F. Schisterman, and A.~I. Naimi,
  ``The role of the natural course in causal analysis,'' {\em American Journal
  of Epidemiology}, vol.~191, no.~2, pp.~341--348, 2022.

\bibitem{Rudolph2018}
J.~E. Rudolph, S.~R. Cole, and J.~K. Edwards, ``Parametric assumptions equate
  to hidden observations: comparing the efficiency of nonparametric and
  parametric models for estimating time to aids or death in a cohort of
  hiv-positive women,'' {\em BMC Medical Research Methodology}, vol.~18, no.~1,
  pp.~1--5, 2018.

\bibitem{Chernozhukov2018}
V.~Chernozhukov, D.~Chetverikov, M.~Demirer, E.~Duflo, C.~Hansen, W.~Newey, and
  J.~Robins, ``Double/debiased machine learning for treatment and structural
  parameters,'' 2018.

\bibitem{Naimi2021ML}
A.~Naimi, A.~Mishler, and E.~Kennedy, ``Challenges in obtaining valid causal
  effect estimates with machine learning algorithms.,'' {\em American Journal
  of Epidemiology}, 2021.

\bibitem{Zivich2021}
P.~N. Zivich and A.~Breskin, ``Machine learning for causal inference: on the
  use of cross-fit estimators,'' {\em Epidemiology (Cambridge, Mass.)},
  vol.~32, no.~3, pp.~393--401, 2021.

\bibitem{Westreich2010PS}
D.~Westreich, J.~Lessler, and M.~J. Funk, ``Propensity score estimation: neural
  networks, support vector machines, decision trees (cart), and
  meta-classifiers as alternatives to logistic regression,'' {\em Journal of
  Clinical Epidemiology}, vol.~63, no.~8, pp.~826--833, 2010.

\bibitem{Lehmann1999}
E.~L. Lehmann, {\em Elements of large-sample theory}.
\newblock Springer, 1999.

\bibitem{Lesko2017}
C.~R. Lesko, A.~L. Buchanan, D.~Westreich, J.~K. Edwards, M.~G. Hudgens, and
  S.~R. Cole, ``Generalizing study results: a potential outcomes perspective,''
  {\em Epidemiology (Cambridge, Mass.)}, vol.~28, no.~4, p.~553, 2017.

\bibitem{Sun2018}
B.~Sun, N.~J. Perkins, S.~R. Cole, O.~Harel, E.~M. Mitchell, E.~F. Schisterman,
  and E.~J. Tchetgen~Tchetgen, ``Inverse-probability-weighted estimation for
  monotone and nonmonotone missing data,'' {\em American Journal of
  Epidemiology}, vol.~187, no.~3, pp.~585--591, 2018.

\bibitem{Sun2018a}
B.~Sun and E.~J. Tchetgen~Tchetgen, ``On inverse probability weighting for
  nonmonotone missing at random data,'' {\em Journal of the American
  Statistical Association}, vol.~113, no.~521, pp.~369--379, 2018.

\bibitem{Robins2000a}
J.~M. Robins and D.~M. Finkelstein, ``Correcting for noncompliance and
  dependent censoring in an aids clinical trial with inverse probability of
  censoring weighted (ipcw) log-rank tests,'' {\em Biometrics}, vol.~56, no.~3,
  pp.~779--788, 2000.

\bibitem{Howe2016}
C.~J. Howe, S.~R. Cole, B.~Lau, S.~Napravnik, and J.~J. Eron~Jr, ``Selection
  bias due to loss to follow up in cohort studies,'' {\em Epidemiology
  (Cambridge, Mass.)}, vol.~27, no.~1, p.~91, 2016.

\bibitem{Edwards2015}
J.~K. Edwards, S.~R. Cole, and D.~Westreich, ``All your data are always
  missing: incorporating bias due to measurement error into the potential
  outcomes framework,'' {\em International Journal of Epidemiology}, vol.~44,
  no.~4, pp.~1452--1459, 2015.

\bibitem{Westreich2017c}
D.~Westreich, J.~K. Edwards, C.~R. Lesko, E.~Stuart, and S.~R. Cole,
  ``Transportability of trial results using inverse odds of sampling weights,''
  {\em American Journal of Epidemiology}, vol.~186, no.~8, pp.~1010--1014,
  2017.

\end{thebibliography}
\bibliographystyle{ieeetr}

\end{document}